\documentclass[aps,pra,twocolumn,superscriptaddress,showpacs]{revtex4}
\usepackage{bm}
\usepackage{graphicx,color}
\usepackage{amsmath}
\usepackage{amssymb}
\usepackage{amsfonts}
\usepackage{latexsym}

\newcommand{\Tr}{\mathop{\text{Tr}}\nolimits}
\newcommand{\ket}[1]{|{#1}\rangle}
\newcommand{\bra}[1]{\langle{#1}|}

\newcommand{\ketbras}[3]{\ket{#1}_{#3}\hspace*{-0.2mm}\bra{#2}}

\definecolor{dgreen}{rgb}{0,0.5,0}

\definecolor{delete}{cmyk}{0.5,0,0,0}

\begin{document}
\title{State Tomography of a Qubit through Scattering of a Probe Qubit}
\author{Antonella De Pasquale}
\affiliation{Dipartimento di Fisica, Universit\`a di Bari, I-70126 Bari, Italy}
\affiliation{Istituto Nazionale di Fisica Nucleare, Sezione di Bari, I-70126 Bari, Italy}
\affiliation{MECENAS, Universit\`a Federico II di Napoli, Via Mezzocannone 8, I-80134 Napoli, Italy}
\author{Kazuya Yuasa}
\affiliation{Waseda Institute for Advanced Study, Waseda University, Tokyo 169-8050, Japan}
\author{Hiromichi Nakazato}
\affiliation{Department of Physics, Waseda University, Tokyo 169-8555, Japan}

\date[]{July 30, 2009}

\begin{abstract}
We discuss the state tomography of a fixed qubit (a spin-1/2 target particle), which is in general in a mixed state, through 1D scattering of a probe qubit off the target.
Two strategies are presented, by making use of different degrees of freedom of the probe, spin and momentum.
Remarkably, the spatial degree of freedom of the probe can be useful for the tomography of the qubit.
\end{abstract}

\maketitle

\section{Introduction}
\label{sec:Introduction}
The determination of a quantum state is a highly nontrivial problem.
A wave function itself, or more generally a density operator, is not an observable and cannot be measured directly.
One can see the state only through measurable quantities, which are related to the matrix elements of the density operator.
It is therefore an interesting and important issue to discuss how to infer a given quantum state from a list of observed quantities.
Such a problem is called state reconstruction or state tomography \cite{ref:StateEstimation}.

Recent advances in technology have been enabling us to engineer
variety of peculiar quantum states to explore fundamental aspects of
quantum mechanics. Generation of highly quantum states is one of the
key elements for the realization of the ideas of quantum information
\cite{ref:NielsenChuang}. It is clear that such issues cannot go
without supports by the state reconstruction technology. Probing
nontrivial correlations in a quantum system also helps us to clarify
intrinsic characters of various many-body systems.

From a practical point of view, direct accesses to target systems
are often limited. In such a case, we have to explore a way to probe
the target in an indirect way. Scattering has always been considered
a very powerful way to investigate many physical systems in a wide
range of fields of physics, from elementary-particle physics to
condensed-matter physics. Loosely speaking, all the physical
processes to access targets can be regarded as scattering processes.

In this paper, we focus on the state tomography of a qubit (spin-1/2 particle) via scattering, in a simple 1D setup.
We send a probe qubit to the fixed target qubit and see the state of the probe after its scattering off the target.
Its scattering data contain the information on the target state before the scattering, from which we reconstruct the state.
In order to reconstruct the qubit state, three independent scattering data are required.
One possibility is to make use of the spin degree of freedom of the probe.
A collection of three transmission/reflection probabilities with three different sets of the initial and final spin states of the probe provides sufficient information for the reconstruction of the target qubit state.

Notice here that in our setup, there is an additional (spatial) degree of freedom available in the scheme for the tomography of the spin, i.e., the momentum of the probe.
It will be shown that this degree of freedom can be utilized to ``optimize" (in the sense prescribed below) the scheme.
It is interesting to see that the \textit{spatial} degree of freedom can play a central role for the tomography of the \textit{spin}.
It will be demonstrated that three scattering data required for the qubit tomography are  available by arranging different incident momenta and scattering directions, with the initial and final spin states of the probe being fixed.

\section{Setup}
\label{scattering problem}
\begin{figure}[t]
\includegraphics[width=0.48\textwidth]{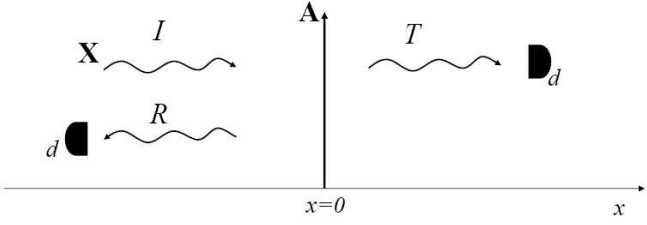}
\caption{A probe qubit $X$ is sent with a fixed wave number $k$ from
the left to a target qubit $A$ fixed at $x=0$, scattered by the
delta-shaped potential produced by the target qubit $A$ with a
spin-spin interaction of the Heisenberg type, and detected on the
right or on the left with spin-sensitive detectors.}
\label{fig:scattering picture}
\end{figure}

Suppose that a qubit $A$ is fixed at $x=0$ on a 1D line.
We are going to discuss the reconstruction of its state, which is in general a mixed state $\rho_A$, through the scattering data of a probe qubit $X$ off the target $A$\@.
We assume that the following Hamiltonian describes the scattering process:
\begin{equation}
H=\frac{p^2}{2m}+g(\bm{\sigma}_X\cdot\bm{\sigma}_A)\delta{(x)}.
\label{eqn:Hamiltonian}
\end{equation}
Here $x$ and $p\,(=-i \hbar\, d/dx)$ are respectively the position
and the momentum of the probe qubit $X$ of mass $m$,
and the potential produced by the fixed qubit $A$ is assumed to be
represented by the delta function. When $X$ is scattered by $A$,
their spins interact with each other through the Heisenberg-type
interaction with a positive coupling constant $g$, with
$\bm{\sigma}_J$ being the Pauli operators acting on qubit
$J\,(=X,A)$. See Fig.\ \ref{fig:scattering picture}.

Let us start by solving the scattering problem of the Hamiltonian
(\ref{eqn:Hamiltonian}). The probe $X$ is sent from the left ($x<0$)
to the target $A$ with a fixed incident wave number $k\,(>0)$, and
scattered.
The scattering (S) matrix element is given by \cite{ref:qpfesc}
\begin{align}
&\langle k'\zeta'|S|k\zeta\rangle
\nonumber\\
&\quad\ \ %
=\delta(k'-k)\delta_{\zeta'\zeta}
-2\pi i\delta(E_{k'}-E_k)\langle k'\zeta'|V|\Psi_k\zeta\rangle,
\end{align}
where $|k\zeta\rangle$ is the eigenstate of the free Hamiltonian $H_0=p^2/2m$ and $|\Psi_k\zeta\rangle$ that of the total Hamiltonian $H=H_0+V$, both belonging to the same eigenvalue $E_k=\hbar^2k^2/2m$, with $\zeta$ denoting the spin degrees of freedom of two qubits.
The coordinate representation of the latter reads as \cite{ref:qpfesc}
\begin{align}
&\langle x|\Psi_k\zeta\rangle\nonumber\\
&\quad\ %
=\langle x|k\zeta\rangle-\int dx'\,G_k(x-x')\frac{2m}{\hbar^2}V(x')\langle x'|\Psi_k\zeta\rangle,
\end{align}
where $G_k(x)=i(2k)^{-1}e^{ik|x|}$ is the retarded Green function in 1D\@.
It is easy to calculate the source term at $x'=0$,
\begin{align}
\langle0|\Psi_k\zeta\rangle&=\langle0|k\zeta\rangle-i\Omega
({\bm\sigma}_X\cdot{\bm\sigma}_A)
\langle0|\Psi_k\zeta\rangle\nonumber\\
&=(1+i\Omega{\bm\sigma}_X\cdot{\bm\sigma}_A)^{-1}\langle0|k\zeta\rangle,
\end{align}
with a dimensionless parameter
\begin{equation}
\Omega=\frac{mg}{\hbar^2k}.
\end{equation}
Recall that the Heisenberg-type coupling $\bm\sigma_X\cdot\bm\sigma_A$ is rewritten as
\begin{equation}
\bm\sigma_X\cdot\bm\sigma_A=-3{\cal P}_1+{\cal P}_3
\end{equation}
in terms of the projection operators
\begin{equation}
{\cal P}_1=\frac{1-\bm\sigma_X\cdot\bm\sigma_A}{4},\quad
{\cal P}_3=\frac{3+\bm\sigma_X\cdot\bm\sigma_A}{4},
\end{equation}
onto the spin-singlet and -triplet eigenspaces, respectively.
This allows us to evaluate the inverse operator as
\begin{equation}
\frac{1}{1+i\Omega\bm\sigma_X\cdot\bm\sigma_A}=\frac{1}{1-3i\Omega}{\cal P}_1+\frac{1}{1+i\Omega}{\cal P}_3.
\end{equation}
We therefore obtain
\begin{equation}
\langle k'\zeta'|S|k\zeta\rangle\nonumber\\
=\delta(k'-k)\langle\zeta'|T|\zeta\rangle+\delta(k'+k)\langle\zeta'|R|\zeta\rangle,
\label{eq:Smtrxsdots}
\end{equation}
where the scattering matrices responsible for transmission and reflection, $T$ and $R$, read
\begin{align}
R&=r_1{\cal P}_1+r_3{\cal P}_3\nonumber\\
 &=\frac{1}{(1-3i\Omega)(1+i\Omega)}(-3\Omega^2-i\Omega\bm\sigma_X\cdot\bm\sigma_A),\\
T&=1+R=t_1{\cal P}_1+t_3{\cal P}_3\nonumber\\
 &=\frac{1}{(1-3i\Omega)(1+i\Omega)}(1-2i\Omega-i\Omega\bm\sigma_X\cdot\bm\sigma_A).
\end{align}
Here the coefficients $r_{1(3)}$ and $t_{1(3)}$ coincide with the reflection and transmission amplitudes calculated separately for the spin-singlet(triplet) eigenspace, in which the interaction Hamiltonian is given by a scalar (i.e., not spin-dependent) potential $-3g\delta(x)$ for the spin-singlet case or by $g\delta(x)$ for the spin-triplet case.
This implies that the problem can be reduced to an ordinary scattering problem of a spinless particle.

\subsection{Scattering probabilities}\label{scattering probabilities}
\begin{figure}
\begin{tabular}{r}
\includegraphics[height=50truemm]{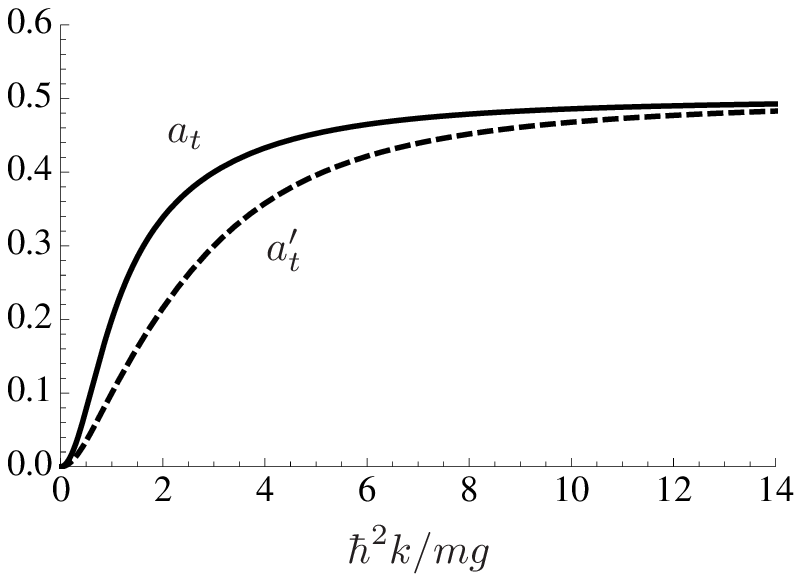}\\
\includegraphics[height=50truemm]{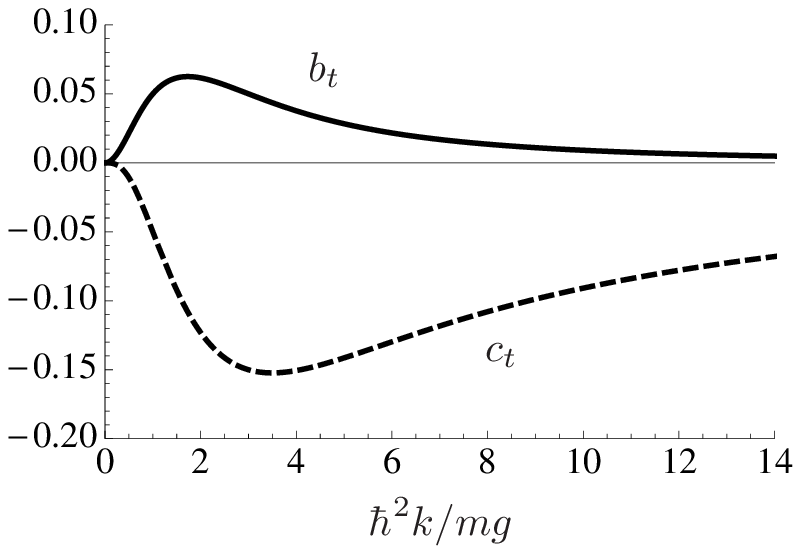}
\end{tabular}
\caption{
The coefficients in the transmission probability $P^t$ in
(\ref{transmission_probability})--(\ref{eqn:ct}), as functions of
the incident wave number $k$.}\label{fig_transmission_coeff}
\end{figure}

\begin{figure}
\begin{tabular}{r}
\includegraphics[height=50truemm]{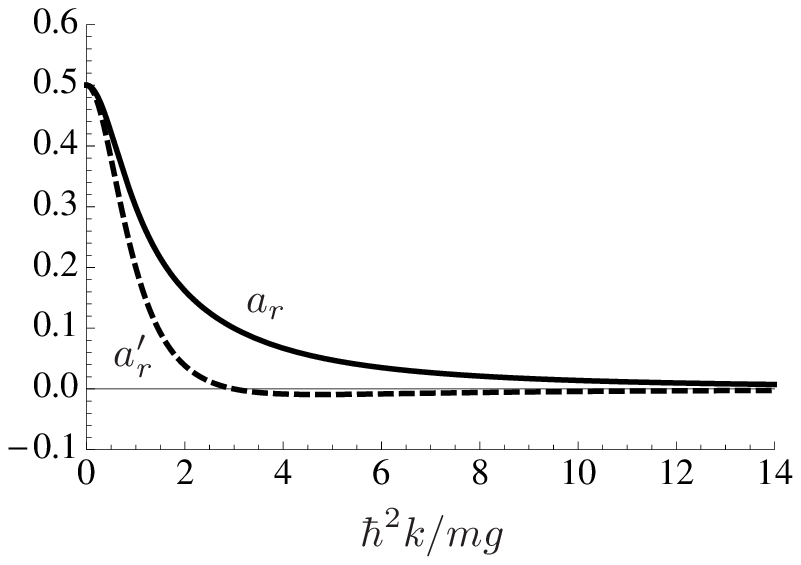}\\
\includegraphics[height=50truemm]{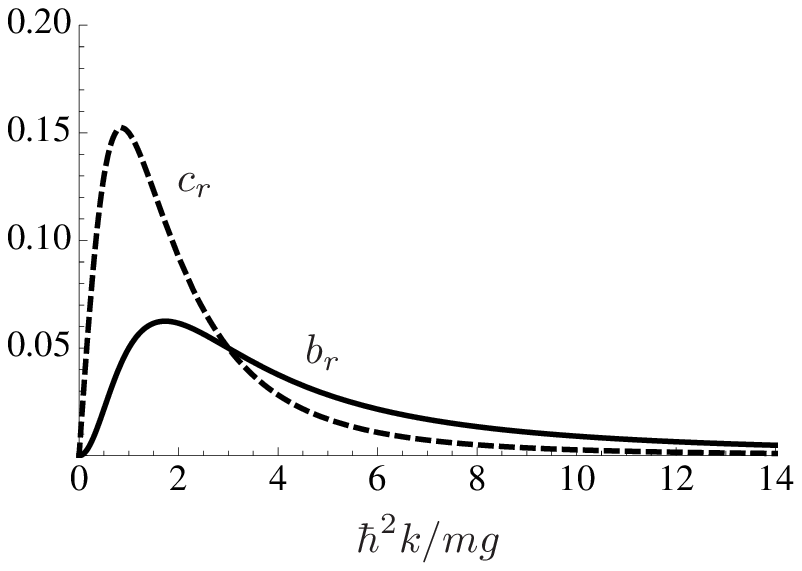}
\end{tabular}
\caption{
The coefficients in the reflection probability $P^r$ in
(\ref{reflection_probability})--(\ref{eqn:cr}), as functions of the
incident wave number $k$.}\label{fig_reflection_coeff}
\end{figure}

In the tomographic schemes we are going to discuss in the following, the probe $X$ is sent with its spin polarized to the direction specified by a unit vector ${\bm n}_i$, and we see the probability of $X$ being transmitted or reflected with its spin rotated to ${\bm n}_f$.
Such probabilities are given by
\begin{equation}
P=\Tr\{
\ketbras{\psi_f}{\psi_f}{X}S(\ketbras{\psi_i}{\psi_i}{X}\otimes\rho_A) S^\dag\},
\end{equation}
where $\ket{\psi_i}_X$ and $\ket{\psi_f}_X$ are the incident and final spin states of $X$, which are expressed in the Bloch-sphere formalism as
\begin{equation}
\ketbras{\psi_i}{\psi_i}{X}=\frac{1+{\bm n}_i\cdot\bm{\sigma}_{X}}{2},\quad
\ketbras{\psi_f}{\psi_f}{X}=\frac{1+{\bm n}_f\cdot\bm{\sigma}_X}{2},
\end{equation}
and $S$ is to be substituted by $T$ or $R$ depending on whether $X$ is transmitted or reflected.
The target qubit $A$ is in general in a mixed state $\rho_A$, which is characterized by a Bloch vector $\bm{v}$ such that
\begin{equation}\label{target initial state}
\rho_A=\frac{1+\bm{v}\cdot\bm{\sigma}_A}{2}\quad
(0\le|\bm{v}|\le1).
\end{equation}
The transmission probability for a given set $\{{\bm n}_i,{\bm n}_f,k\}$ reads as
\begin{align}\label{transmission_probability}
P^t
&=a_t(\Omega)
+ a'_t(\Omega)({\bm n}_f\cdot{\bm n}_i)
\nonumber\\
&\quad
+ b_t(\Omega)[3({\bm n}_f\cdot\bm{v})
+
({\bm n}_i\cdot\bm{v})]
+c_t(\Omega)\bm{v}\cdot({\bm n}_f\times{\bm n}_i)
\end{align}
with
\begin{align}
a_t(\Omega)&=\frac{1+7\Omega^2}{2(1+\Omega^2)(1+9\Omega^2)}, \label{at}\\
a'_t(\Omega)&=\frac{1+3\Omega^2}{2(1+\Omega^2)(1+9\Omega^2)},\label{aPrimet}\\
b_t(\Omega)&=\frac{\Omega^2}{(1+\Omega^2)(1+9\Omega^2)},\label{bt}
\\c_t(\Omega)&=-\frac{\Omega}{(1+\Omega^2)(1+9\Omega^2)}.
\label{eqn:ct}
\end{align}
These coefficients are plotted in Fig.\ \ref{fig_transmission_coeff} as functions of the incident wave number $k$ of the probe qubit $X$.
Similarly, the probability for the reflected case reads
\begin{align}\label{reflection_probability}
P^r
&=a_r(\Omega) + a'_r(\Omega)
({\bm n}_f\cdot{\bm n}_i)
\nonumber\\
&
\quad
+ b_r(\Omega) [({\bm n}_f\cdot\bm{v})-
({\bm n}_i\cdot\bm{v})]+c_r(\Omega) \bm{v}\cdot({\bm n}_f\times{\bm n}_i)\end{align}
with
\begin{align}
a_r(\Omega)&=\frac{3\Omega^2(1+3\Omega^2)}{2(1+\Omega^2)(1+9\Omega^2)}, \\
a'_r(\Omega)&=-\frac{\Omega^2(1-9\Omega^2)}{2(1+\Omega^2)(1+9\Omega^2)},\\
b_r(\Omega)&=\frac{\Omega^2}{(1+\Omega^2)(1+9\Omega^2)},\label{br}\\
c_r(\Omega)&=\frac{3\Omega^3}{(1+\Omega^2)(1+9\Omega^2)},
\label{eqn:cr}
\end{align}
which are plotted in Fig.\ \ref{fig_reflection_coeff}.
Observe that the transmission and reflection probabilities $P^t$ and $P^r$ are both spherically symmetric in spin space.
This reflects the symmetry of the Heisenberg coupling $\bm{\sigma}_X\cdot\bm{\sigma}_A$:
no preferred direction is present in the system.

\section{Strategies for the State Tomography}
Let us now discuss the tomography of the state $\rho_A$, through the
scattering data $P^{t,r}$. Our objective is to determine the three
independent components of vector $\bm{v}$, which exactly corresponds
to the complete specification of the state $\rho_A$, see
(\ref{target initial state}). From an experimental point of view,
this means that we need to arrange three independent experimental
setups. As is clear from (\ref{transmission_probability}) and
(\ref{reflection_probability}), the scattering probabilities
$P^{t,r}$ explicitly depend on the initial and detection
orientations ${\bm n}_i$ and ${\bm n}_f$ of the spin of the probe
qubit $X$ as well as on the parameter $\Omega$ containing the
incident wave number $k$ and the coupling constant $g$; we can
consider different strategies for the tomography, by properly
choosing these parameters. In the next subsections, we discuss two
of such tomographic strategies, in which the spatial and spin
degrees of freedom of $X$ play different roles.

In the first approach presented in Sec.\ \ref{first strategy}, a central role is played by the spin degree of freedom of $X$: we tune the incident and detection orientations ${\bm n}_i$ and ${\bm n}_f$ of its spin, while the incident wave number $k$ is fixed.
In order to completely reconstruct $\rho_A$, we arrange
the orientations ${\bm n}_i$ and ${\bm n}_f$ in three independent ways.
For instance, we can choose three different detection orientations ${\bm n}_f$ with the orientation of the incident spin ${\bm n}_i$ being fixed, or we can tune both of them at the same time.
The remaining degree of freedom, the wave vector $k$, also plays an active role: we can ``optimize" in the sense described below the tomographic scheme by tuning it appropriately.

In Sec.\ \ref{par:second strategy}, we discuss another approach,
in which we fully make use of the spatial degree of freedom $k$.
The tomography of the \textit{spin} state $\rho_A$ is possible through the \textit{spatial} degree of freedom.
We collect three independent scattering data for different incident wave numbers $k$ and scattering directions (i.e., transmission or reflection) when the incident and detection orientations of the spin ${\bm n}_i$ and ${\bm n}_f$ are fixed, from which we reconstruct the spin state $\rho_A$.
Furthermore, one can follow the same criterion introduced for the first strategy in order to ``optimize" the tomographic scheme.

\subsection{Strategy I: tuning ${\bm n}_i$ and ${\bm n}_f$ with $k$ being fixed}\label{first strategy}
In this section, we reconstruct the initial state of the target qubit
$A$ by tuning the incident and detection orientations ${\bm n}_i$ and ${\bm n}_f$ of the spin of the probe qubit $X$ with a fixed incident wave number $k$.
In particular, we
discuss two possible examples of this first tomographic strategy.

Let us first fix the orientation of the spin of the incident qubit $X$, ${\bm n}_i$, and choose three different detection orientations ${\bm n}_f=\bm{n}_1,\bm{n}_2,\bm{n}_3$, which can be chosen to be orthogonal to each other.
For instance, choosing $\bm{n}_1$ perpendicular to ${\bm n}_i$,
\begin{equation}
\bm{n}_1,\quad
\bm{n}_2=\bm{n}_i\times{\bm n}_1,\quad
\bm{n}_3={\bm n}_i
\label{eq:RefSystemF}
\end{equation}
would be a natural choice as a reference frame.
The transmission probabilities associated to these detection orientations with a fixed wave number $k$ read
\begin{equation}\label{trasmission_first_strategy}
\begin{pmatrix}
P^t_1 \\
P^t_2 \\
P^t_3
\end{pmatrix}= \begin{pmatrix}3 b_t & -c_t &b_t \\
c_t & 3 b_t &b_t \\
0 & 0 & 4 b_t \end{pmatrix}\begin{pmatrix}
v_1\\
v_2\\
v_3
\end{pmatrix}+\begin{pmatrix}a_t  \\
a_t \\
a_t+a'_t
\end{pmatrix},
\end{equation}
where
\begin{equation}
v_a=\bm{n}_a\cdot\bm{v}
\quad(a=1,2,3)
\end{equation}
are the three independent
components of the target vector $\bm{v}$ along the axes of the reference
system $\{\bm{n}_1,\bm{n}_2,\bm{n}_3\}$ introduced above and $P^t_a$ the transmission probability when the spin of $X$ is detected along direction ${\bm n}_a$.
The three components of the vector $\bm{v}$ are readily obtained by inverting the relation (\ref{trasmission_first_strategy}) as
\begin{equation}\label{v components}
\begin{pmatrix}v_1\\
v_2\\
v_3\end{pmatrix}
=M_t\left[
\begin{pmatrix}
P^t_1\\
P^t_2\\
P^t_3
\end{pmatrix}
-\begin{pmatrix}a_t  \\
a_t \\
a_t+a'_t
\end{pmatrix}
\right]
\end{equation} with
\begin{equation}\label{M_t}
M_t(\Omega)=\frac{1+\Omega^2}{4\Omega^2}\begin{pmatrix}12\Omega^2 & -4\Omega&\Omega(1-3\Omega)\\
4\Omega&12\Omega^2&-\Omega(1+3\Omega)\\
0 & 0& 1+9\Omega^2
\end{pmatrix},
\end{equation}
which completes the reconstruction of the state $\rho_A$\@.
Observe that the only condition needed to be satisfied in order to
invert the matrix in (\ref{trasmission_first_strategy}) to obtain $M_t$ in (\ref{M_t}) is $0<k<\infty$, which can be considered as a self-consistency condition for the present tomographic scheme.

Another example of this first strategy, which is simpler from a
computational point of view, is detecting the spin $X$ oriented in the
same direction as the incident spin, ${\bm n}_f={\bm n}_i$.
In this case, the transmission probability reads
\begin{equation}\label{first_strategy f=n}
P^t= a_t + a'_t + 4 b_t ({\bm n}_i\cdot\bm{v}) ,
\end{equation}
from which we immediately obtain
\begin{equation}\label{component of v along n}
{\bm n}_i\cdot\bm{v}= \frac{P^t(1+\Omega^2)(1+9\Omega^2)-(1+5\Omega^2)}{4\Omega^2} .
\end{equation}
By choosing three different orientations for ${\bm n}_i={\bm n}_f$, we gain the complete information on the vector $\bm{v}$, that is, on the state $\rho_A$.
Observe that also in this case the self-consistency condition reads $0<k<\infty$.

Similar schemes are available with reflection probabilities
$P^r$. Notice however that, if we choose ${\bm n}_f={\bm n}_i$ as
we did in the above two schemes, no information on $\bm{v}$ is
attainable from the reflection probability $P^r$,
see (\ref{reflection_probability}), where $\bm{v}$ disappears
from $P^r$ for ${\bm n}_f={\bm n}_i$.
A possible solution to this problem is to flip the orientations $\bm {n}_a\to-\bm{n}_a\,(a=1,2,3)$ in the choice of the reference system
$\{\bm{n}_1,\bm{n}_2,\bm{n}_3\}$ in (\ref{eq:RefSystemF}) for the former scheme, while measuring the reflected $X$ in $-{\bm n}_i$ direction instead of ${\bm n}_i$ for the latter.
In this way, the schemes with the reflection probabilities work
similarly to the ones with the transmission probabilities, with the same self-consistency condition needed for the tomographic reconstruction of the initial state of the target qubit, $\rho_A$.

The reason why these tomographic schemes that make use of the spin degree of freedom of $X$ work is the following.
Let $\ket{{\uparrow}({\downarrow})}$ denote the state with spin parallel (anti-parallel) to the orientation ${\bm n}_i$ of the incident probe spin $X$.
Then, the state of $A$ is in general expressed as
\begin{equation}
\rho_A= \rho_{\uparrow\uparrow}\ketbras{\uparrow}{\uparrow}{A}
+\rho_{\downarrow\downarrow}\ketbras{\downarrow}{\downarrow}{A}
+\rho_{\uparrow\downarrow}\ketbras{\uparrow}{\downarrow}{A}
+\rho_{\downarrow\uparrow}\ketbras{\downarrow}{\uparrow}{A},
\end{equation}
and the spin state of the probe $X$ after the transmission by $A$ becomes (apart from the normalization)
\begin{align}
\tilde{\rho}_X
\propto&\Tr_A\Bigl\{
T\,\Bigl(
\ketbras{\uparrow}{\uparrow}{\text{X}}\otimes\rho_A
\Bigr)\,T^\dag
\Bigr\}
\nonumber\\
\propto&
\left(
\rho_{\uparrow\uparrow}
+\frac{1+\Omega^2}{1+9\Omega^2}
\rho_{\downarrow\downarrow}
\right)\ketbras{\uparrow}{\uparrow}{X}
+\frac{4\Omega^2}{1+9\Omega^2}
\rho_{\downarrow\downarrow}
\ketbras{\downarrow}{\downarrow}{X}
\nonumber\\
&
{}+\frac{2i\Omega}{1+3i\Omega}
\rho_{\uparrow\downarrow}
\ketbras{\uparrow}{\downarrow}{X}
-\frac{2i\Omega}{1-3i\Omega}
\rho_{\downarrow\uparrow}
\ketbras{\downarrow}{\uparrow}{X}.
\end{align}
A similar expression is available for the reflection case.
Observe that the component $\rho_{\uparrow\uparrow}$ of $\rho_A$ is
associated to $\ketbras{\uparrow}{\uparrow}{X}$ in $\tilde{\rho}_X$,
$\rho_{\uparrow\downarrow}$ to $\ketbras{\uparrow}{\downarrow}{X}$,
and so on: the spin state $\rho_A$ is more or less ``transferred''
to $X$ after the scattering. This is due to the Heisenberg coupling
$\bm{\sigma}_X\cdot\bm{\sigma}_A$, which ``swaps'' the states
between $A$ and $X$\@. This is why we can see the spin state of $A$
by looking at the spin state of $X$.

\begin{figure}[t]
\includegraphics[height=53truemm]{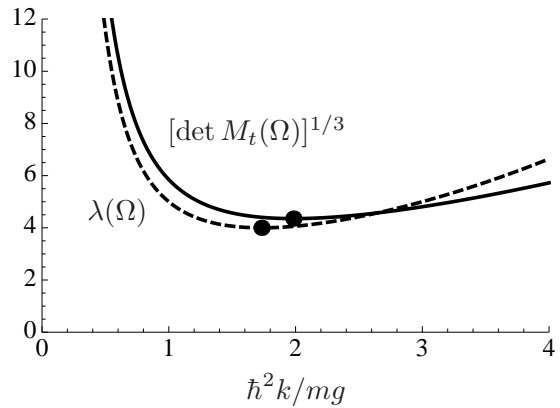}
\caption{
The Jacobian of the linear map (\ref{v components}) between the probability space and the Bloch sphere, $\det M_t(\Omega)$ given in (\ref{eqn:DetM}) (solid), and the coefficient $\lambda(\Omega)$ (dashed), as functions of the incident wave number $k$ of the probe $X$, which are both to be minimized by tuning $k$ for the optimizations of the two approaches of Strategy I, presented in the text.
$[\det M_t(\Omega)]^{1/3}$ is actually plotted instead of $\det M_t(\Omega)$, to better compare it with $\lambda(\Omega)$.
$[\det M_t(\Omega)]^{1/3}$ is minimum at $\hbar^2k/mg=\sqrt{1+\sqrt{127}}/2$, while  $\lambda(\Omega)$ at $\sqrt{3}$ (indicated by dots).}
\label{fig:first strategy optimization}
\end{figure}

Until now, we have shown how the initial state of the target qubit
$A$ can be reconstructed by sending a probe qubit $X$ with a fixed
wave numebr $k$. A natural question would then arise: can we use the
spatial degree of freedom of $X$ to optimize the above tomographic
schemes in the sense that possible errors in the scattering data can
least affect the determination of the state? It is actually
possible. For instance, one would be able to reduce the effects of
possible errors in the observations of the probabilities $P^{t,r}$
on the reconstructed vector $\bm{v}$, by properly tuning the
incident wave number $k$. Observe that (\ref{v components}) and
(\ref{component of v along n}) are linear mappings between the Bloch
sphere of radius $1$ and the probability space, associated to the
scattering data $P^{t,r}$. The reconstructed vector $\bm{v}$ is
least sensitive to the errors in the observed probabilities
$P^{t,r}$, when the volume of this probability space is maximum.
Stated differently, under such a condition, the probabilities
$P^{t,r}$ are the most sensitive to the Bloch vector $\bm{v}$ to be
reconstructed, and one can do a better tomography. For the first
scheme, in which we tune the detection orientation ${\bm n}_f$ of
the spin of the probe $X$, the volume associated to the
probabilities is maximum when the Jacobian of the map (\ref{v
components}), which is given by the determinant
\begin{equation}
\det M_t(\Omega)= \frac{(1+\Omega^2)^3 (1+9
\Omega^2)^2}{4 \Omega^4},
\label{eqn:DetM}
\end{equation}
is minimum.
For the second scheme, in which the incident and detection orientations of the probe spin $X$ are the same ${\bm n}_f={\bm n}_i$, the coefficient multiplying the scattering
probability $P^t$ in (\ref{component of v along n}),
\begin{equation}
\lambda(\Omega)\equiv\frac{(1+\Omega^2)(1+9\Omega^2)}{4 \Omega^2},
\end{equation}
is to be minimized.
In Fig.\ \ref{fig:first strategy optimization}, $[\det M_t(\Omega)]^{1/3}$ and $\lambda(\Omega)$ are plotted as functions of the incident wave number $k$ of
the probe $X$.
They become minimum at $1/\Omega=\hbar^2k/mg=\sqrt{1+\sqrt{217}}/2 \simeq 1.98$ and $\sqrt{3}$, respectively.
In particular, observe that, at $\hbar^2k/mg=\sqrt{3}$, the formula for the tomography (\ref{component of v along n}) is reduced to
\begin{equation}
{\bm n}_i\cdot\bm{v}= 4 P^t -2,
\end{equation}
from which it immediately follows that the probability $P^t$ ranges
\begin{equation}
\frac{1}{4}\leq P^t\leq \frac{3}{4},
\end{equation}
which is the maximum in this scheme.
The same analysis can be applied to the schemes with reflection probabilities, for which the optimal momenta are $\hbar^2k/mg=\sqrt{(-3+\sqrt{33})/2}\simeq1.17$
and $\sqrt{3}$ for the two schemes, respectively.

\subsection{Strategy II: tuning $k$ with ${\bm n}_i$ and ${\bm n}_f$ being fixed}
\label{par:second strategy}
The \textit{spatial} degree of freedom of $X$ can itself play a fundamental role for the
tomographic reconstruction of the target \textit{spin} state $\rho_A$\@.
We fix the incident and detection orientations $\bm{n}_i$ and $\bm{n}_f$ of the qubit $X$ and tune the wave number $k$ to collect sufficient number of scattering data required for the tomography of $\rho_A$\@.

\begin{figure}[b]
\includegraphics[width=0.4\textwidth]{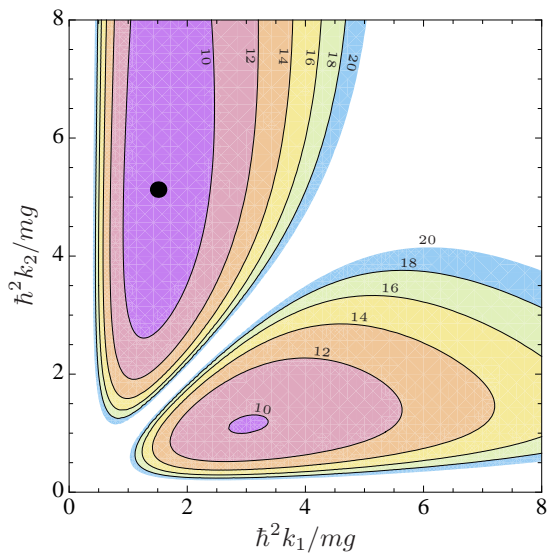}
\caption{(Color online)
Contour plot of $|{\det N(\Omega_1,\Omega_2)}|^{1/3}$ given in
(\ref{determinant__second strategy}), as a function of $k_1$ and
$k_2$.
Only the contours in the range $|{\det N(\Omega_1,\Omega_2)}|^{1/3}\le 20$ are shown.
$|{\det N(\Omega_1,\Omega_2)}|^{1/3}$ takes its minimum value $\simeq8.74$ at $(k_1,k_2)\simeq(1.51,5.13)$ in the unit $mg/\hbar^2$ (indicated by a dot).}
\label{fig:optimal
momenta second strategy}
\end{figure}
Let us fix ${\bm n}_f$ perpendicular to ${\bm n}_i$ and select three different incident wave numbers $k=k_1,k_2,k_3$.
A collection of the three transmission probabilities with these wave numbers yields
\begin{align}\label{secondStrategy1}
\begin{pmatrix}P^t(\Omega_1) \\
P^t(\Omega_2)  \\
P^t(\Omega_3)\end{pmatrix}
={}&\begin{pmatrix}3 b_t (\Omega_1)&b_t (\Omega_1) &c_t (\Omega_1) \\ 3 b_t (\Omega_2)& b_t (\Omega_2)&c_t (\Omega_2) \\
3 b_t (\Omega_3)& b_t (\Omega_3)& c_t (\Omega_3)
\end{pmatrix}
\begin{pmatrix}
v_f\\
v_i \\
v_\perp
\end{pmatrix}
\nonumber\\
&
{}+\begin{pmatrix}a_t(\Omega_1)
\\
a_t(\Omega_2)
\\
a_t(\Omega_3)
\end{pmatrix},
\end{align}
where $\Omega_j=mg/\hbar^2k_j\,(j=1,2,3)$, and
\begin{equation}
v_f=\bm{n}_f\cdot\bm{v},\quad
v_i=\bm{n}_i\cdot\bm{v},\quad
v_\perp=(\bm{n}_f\times\bm{n}_i)\cdot\bm{v}
\end{equation}
are the three components of the vector $\bm{v}$ with respect to the reference frame fixed by $\{\bm{n}_f,\bm{n}_i,\bm{n}_f\times\bm{n}_i\}$.
This relation cannot be inverted, since two columns of the matrix are proportional to each other, irrespectively of the choice of $\{k_1,k_2,k_3\}$.
This is due to the fact that, once the orientations ${\bm n}_i$ and ${\bm n}_f$ are fixed, only two components of the vector $\bm{v}$ along the directions specified by $3{\bm n}_f+{\bm n}_i$ and ${\bm n}_f\times{\bm n}_i$ are involved in the transmission probability (\ref{transmission_probability}), while it is insensitive to the other component of $\bm{v}$ perpendicular to the plane spanned by these directions.
The same happens if we collect the reflection probabilities (\ref{reflection_probability}) in a similar way.
This problem however can be overcome by combing the transmission and reflection probabilities.
For instance,
\begin{align}\label{good_secondStrategy1}
\begin{pmatrix}
P^r(\Omega_1)  \\
P^t(\Omega_1) \\
P^t(\Omega_2)\end{pmatrix}={}&
\begin{pmatrix}
b_r (\Omega_1)& -b_r (\Omega_1)&c_r(\Omega_1)\\
3 b_t (\Omega_1)&b_t (\Omega_1)&c_t (\Omega_1)\\
3 b_t (\Omega_2)& b_t (\Omega_2)&c_t (\Omega_2)
\end{pmatrix}
\begin{pmatrix}
v_f\\
v_i\\
v_\perp
\end{pmatrix}
\nonumber\\&{} +\begin{pmatrix}
a_r(\Omega_1)\\
a_t(\Omega_1) \\
a_t(\Omega_2)
\end{pmatrix},
\end{align}
which is inverted as
\begin{align}
\begin{pmatrix}
v_f \\
v_i \\
v_\perp
\end{pmatrix}
=N(\Omega_1,\Omega_2)\left[
\begin{pmatrix}
P^r(\Omega_1)  \\
P^t(\Omega_1) \\
P^t(\Omega_2)
\end{pmatrix}
-\begin{pmatrix}
a_r(\Omega_1)\\
a_t(\Omega_1) \\
a_t(\Omega_2)
\end{pmatrix}
\right]
\end{align}
with an inverse matrix $N(\Omega_1,\Omega_2)$, when
\begin{equation}\label{determinant__second strategy}
\det N(\Omega_1,\Omega_2) =-\frac{(1 + {\Omega}_1^2)^2(1+9
{\Omega}_1^2 )^2 (1 + {\Omega}_2^2)(1+9 {\Omega}_2^2 )}{4
{\Omega}_1^3 {\Omega}_2({\Omega}_1 - {\Omega}_2)}
\end{equation}
is nonvanishing and finite.
This condition is fulfilled by $k_1$ and $k_2$ that are both finite and different from zero as well as different from each other.
In this way, we can reconstruct the \textit{spin} state $\rho_A$ by making use of the \textit{spatial} degree of freedom of $X$\@.
Note that only two experimental setups are actually needed to collect the three scattering data: the two probabilities $P^t(\Omega_1)$ and $P^r(\Omega_1)$ are obtained at the same time in a single setup, by sending the probe $X$ with $k_1$ and seeing whether it is transmitted or reflected with its spin orientated to ${\bm n}_f$.

Similarly to the first strategy discussed in the previous subsection, the present scheme is optimized by appropriately choosing the two incident wave numbers $k_1$ and $k_2$ of the probe $X$.
Under the same criterion as the one for the first strategy [maximizing
the volume of the probability space associated by the
linear mapping (\ref{good_secondStrategy1}) to the Bloch sphere of
radius $1$, to which the vector $\bm{v}$ belongs], the optimal choice of $(k_1,k_2)$ is found by minimizing the quantity $|{\det N(\Omega_1,\Omega_2)}|^{1/3}$ from (\ref{determinant__second strategy}).
See Fig.\ \ref{fig:optimal momenta second strategy}, where the optimal set of wave numbers for the present strategy is found at $(k_1,k_2)\simeq(1.51,5.13)$ in the unit $mg/\hbar^2$.

\section{conclusions}
In this paper, we have discussed the state reconstruction/tomography of a fixed qubit through the scattering data of a probe qubit off the target.
We have presented two different strategies for the tomography, in which the spin and spatial degrees of
freedom of the probe qubit play different roles.

The first strategy makes use of the spin degree of freedom of the probe.
The spin state of the target is more or less transferred to the probe spin during the scattering, and therefore, we can infer the spin state of the target through the state tomography of the probe spin.
The other degree of freedom, the momentum of the probe, can be utilized to optimize the tomographic scheme.

The spatial degree of freedom can also play a central role for the state tomography of the target spin.
In the second strategy, three scattering data required for the state tomography of a target spin are collected by choosing different incident wave numbers and scattering directions (transmitted or reflected), with the incident and detection orientations of the probe spin being fixed.
This tomographic scheme can be optimized also by appropriately tuning the set of the incident wave numbers.

The strategies introduced in this paper for a single fixed qubit can be generalized to the tomography of multiple qubits.
In particular, the detection of entanglement would be an important task in the light of quantum information.
In order to reconstruct the state of $N$ spins, we need $4^N-1$ different experimental setups.
Imagine, for instance, how to choose 15 different sets of orientations of the incident and detection spin states of the probe qubit for the two-qubit tomography.
If the number of the target qubits grows, tuning the probe spin to the different orientations required for the first strategy would become more and more difficult.
In such a case, the second strategy may provide a way out of this problem.
It would be worth exploring such a potential of the scheme, which would be an interesting future subject.

\acknowledgments 
This work was done during A.DP.'s stay at Waseda University under the
support by a Special Coordination Fund for Promoting Science and
Technology from the Ministry of Education, Culture, Sports, Science
and Technology, Japan. It is also supported by the bilateral
Italian-Japanese Projects II04C1AF4E on ``Quantum Information,
Computation and Communication'' of the Italian Ministry of
Education, University and Research, by the Joint Italian-Japanese
Laboratory on ``Quantum Information and Computation'' of the Italian
Ministry for Foreign Affairs, by the Grant-in-Aid for Young
Scientists (B) (No.\ 21740294) from the Ministry of Education,
Culture, Sports, Science and Technology, Japan, and by the
Grant-in-Aid for Scientific Research (C) from the Japan Society for
the Promotion of Science.

\end{document}